# Bio-inspired Filter Banks for SSVEP-based Brain-computer Interfaces

A. Fatih Demir, Student Member, IEEE, Huseyin Arslan, Fellow, IEEE, and Ismail Uysal, Member, IEEE

*Abstract—* Brain-computer interfaces (BCI) have the potential to play a vital role in future healthcare technologies by providing an alternative way of communication and control. More specifically, steady-state visual evoked potential (SSVEP) based BCIs have the advantage of higher accuracy and higher information transfer rate (ITR). In order to fully exploit the capabilities of such devices, it is necessary to understand the features of SSVEP and design the system considering its biological characteristics. This paper introduces bio-inspired filter banks (BIFB) for a novel SSVEP frequency detection method. It is known that SSVEP response to a flickering visual stimulus is frequency selective and gets weaker as the frequency of the stimuli increases. In the proposed approach, the gain and bandwidth of the filters are designed and tuned based on these characteristics while also incorporating harmonic SSVEP responses. This method not only improves the accuracy but also increases the available number of commands by allowing the use of stimuli frequencies elicit weak SSVEP responses. The BIFB method achieved reliable performance when tested on datasets available online and compared with two well-known SSVEP frequency detection methods, power spectral density analysis (PSDA) and canonical correlation analysis (CCA). The results show the potential of bio-inspired design which will be extended to include further SSVEP characteristics (e.g. time-domain waveform) for future SSVEP based BCIs.

## I. INTRODUCTION

Brain-computer interface (BCI) research has gained increased attention in recent years due to its great potential to provide an alternate way of communication and control. The number of scientific papers increases exponentially (Fig. 1) and numerous companies directly develop BCIs or aim to integrate BCI-based technology into their product portfolio for future applications [1]. Basically, a BCI is a device that sends user messages or commands to the external world without passing through the brain's normal output pathway of peripheral nerves and muscles [2]. The intended messages can be conveyed through a BCI speller device, or intended user commands can be used to control physical devices, such as a wheelchair or a mobile device (Fig. 2). Although primary target of BCI technology is people with severe neuromuscular disorders now, advanced BCI systems will serve healthy people by providing convenient speed and accuracy in future. Hence, we could eventually be operating devices purely by thinking.

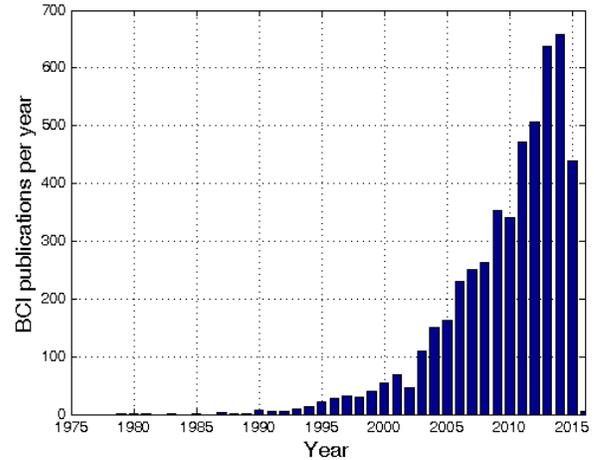

Fig. 1. Number of BCI publications per year (Data is obtained from PubMed on 05 December 2015).

Although there exist several methods to measure the brain activity, electroencephalography (EEG) is widely used in noninvasive BCI applications because of its high time resolution, small size, and inexpensive equipment [3]. Especially, high time resolution is crucial for BCIs to work as real-time systems. Many EEG activities could serve to drive BCIs but steady state evoked visual potential (SSVEP) based BCIs have the advantage of having higher accuracy and higher information transfer rate (ITR) compared to other BCI modalities [4].SSVEPs are stable voltage oscillations in the brain that are elicited by rapid repetitive visual stimulation [5]. A flickering or moving visual stimulus at a constant frequency elicits a response in the occipital region at the same frequency and its harmonics.

This paper introduces bio-inspired filter banks (BIFB) for a novel SSVEP recognition method. It is known that SSVEP response to a flickering visual stimulus is frequency selective and gets weaker as the frequency of the stimuli increases [6,7]. In the proposed approach, the gain and bandwidth of the filters are designed and tuned based on these characteristics while also incorporating harmonic SSVEP responses. The BIFB method is tested on datasets available online and compared with two well-known SSVEP frequency detection methods, power spectral density analysis (PSDA) and canonical correlation analysis (CCA).

A. Fatih Demir, Huseyin Arslan, and Ismail Uysal are with the Department of Electrical Engineering, University of South Florida, USA (e-mail: afdemir @mail.usf.edu, {arslan, iuysal}@usf.edu).

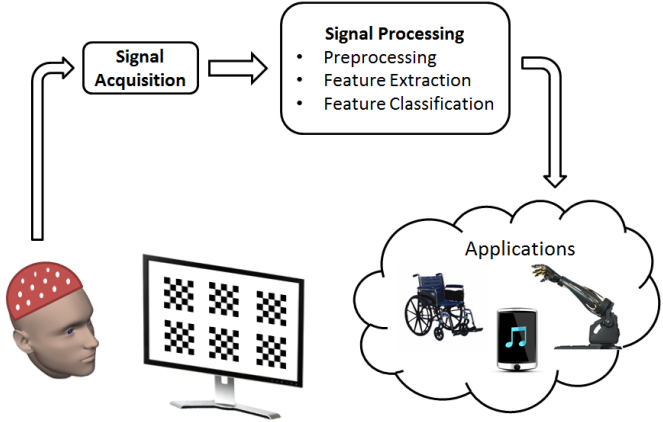

Fig. 2. Functional model of a SSVEP-BCI.

## II. METHODS AND MATERIALS

### A. SSVEP Datasets

In order to test the proposed BIFB method, two datasets available online are used in this study. The first one (AVI [8]) consists of SSVEP experiments done with four test subjects where flickering images at seven different frequencies (6 Hz, 6.5 Hz, 7 Hz, 7.5 Hz, 8.2 Hz, 9.3 Hz, and 10 Hz) are used for visual stimuli. Each trial in a session lasts for 30 seconds, and the trials are repeated at least three times for each frequency. The signal electrode is placed in the occipital channel Oz, using the standard 10-20 system for electrode placement. The second dataset (RIKEN-LABSP [9]) consists of SSVEP experiments conducted with four test subjects where checkerboard pattern reversal stimulation at three frequencies (8 Hz, 14 Hz, and 28 Hz) is used. Each trial in a session lasts for 15 seconds, and the trials are repeated five times for each frequency. The EEG signals are recorded over 128 channels using the ABC layout standard for electrode placement. An overview of these datasets is summarized in Table I and the reader is referred to individual articles for a more detailed description.

### B. SSVEP Frequency Detection using BIFB

In this work, we propose the use of BIFBs for SSVEP frequency detection. It is known that SSVEP response to a flickering visual stimulus is frequency selective and gets weaker as the frequency of the stimuli increases. The gain and bandwidth of the filters are designed and tuned based on these SSVEP characteristics in the proposed approach. Moreover, harmonic frequency responses are also captured by the filter banks using appropriate weights. The design of the BIFB is as follows: assuming there are $K$ target stimuli frequencies $f_1, f_2, \ldots f_K$ in the BCI system, the triangular filters ($H_k$) are expressed by the following equation:

$$H_k(f): \begin{cases} \dfrac{f - (f_k - BW_k/2)}{BW_k} * g_k, & (f_k - BW_k/2) \leq f \leq f_k \\ \dfrac{(f_k + BW_k/2) - f}{BW_k} * g_k, & f_k \leq f \leq (f_k + BW_k/2) \\ 0, & otherwise \end{cases}$$

where $BW_k$ and $g_k$ represent the bandwidth and gain of the filter, respectively. Initially, higher bandwidth and gain are set to frequencies with low SSVEP response [7]. Subsequently, these parameters are optimized for individual users in order to counter frequency selective nature of SSVEP response. The shape of the triangular waveform in the filter banks is chosen to emphasize the center frequency without completely omitting the adjacent frequencies. Fig.3 presents BIFB design for the first dataset and reveals frequency selective nature of the SSVEP response. The second filter bank in Fig. 4 designed for RIKEN-LABSP dataset deals with the weakening of SSVEP response as the frequency increases.

Once the BIFB parameters are trained, the EEG signal from the occipital channel Oz, where the SSVEP response is powerful, is first band-pass filtered (5-35Hz) to reduce noise and normalized. The resulting signal is then analyzed using a time window of length 4s moving with 1s displacement. A Hamming window of the same length is applied before Fast Fourier Transform (FFT) operation to decrease large side lobes. Finally, the power spectrum is estimated by multiplying each signal's FFT with the BIFB to obtain the class value ($c_k$) for each target frequency and ultimately the frequency with maximum class value is determined as follows:

$$c_k(t) = \sum_f |Y| * H_k + \sum_f |Y| * H_{2k} * w_h$$

$$f_t = \max_k(c_k)$$

where $|Y|$ represents power amplitudes of the EEG signal and $w_h$ denotes harmonic frequency weight. Harmonic weights provide to tune frequency selective and subject specific harmonic frequency response. The SSVEP frequency is labeled as detected when the same frequency ($f_t$) occurs at least three times in the last four iterations. If the selection criteria is not satisfied during the given time period, it is evaluated as an unsuccessful detection.

### C. Comparison Methods

In this study, two well-known SSVEP frequency detection methods, PSDA and CCA are implemented as baseline to compare with the proposed BIFB approach. The EEG signal from the occipital channel Oz is similarly preprocessed for PSDA by applying a bandpass filter (5-35Hz), normalizing and passing through a Hamming window of similar length. Then, FFT is performed using the predetermined window length which serves to improve the accuracy after a training session. Peak detection is applied to the windowed data in frequency domain. The detection is classified as successful if the fundamental or second harmonic stimuli frequency is detected as the peak frequency in the performance evaluation.

TABLE I. SSVEP DATASETS TESTED IN THIS STUDY

| Dataset | # of Subjects | # of Trials | Record Length | # of Channels | # of Stimuli Frequencies | Stimuli Frequencies |
|---|---|---|---|---|---|---|
| AVI [8] | 4 | 92 | 30s | 1 | 7 | 6 Hz, 6.5 Hz, 7 Hz, 7.5 Hz, 8.2 Hz, 9.3 Hz, 10 Hz |
| RIKEN LABSP [9] | 4 | 60 | 15s | 128 | 3 | 8Hz, 14Hz, 28Hz |

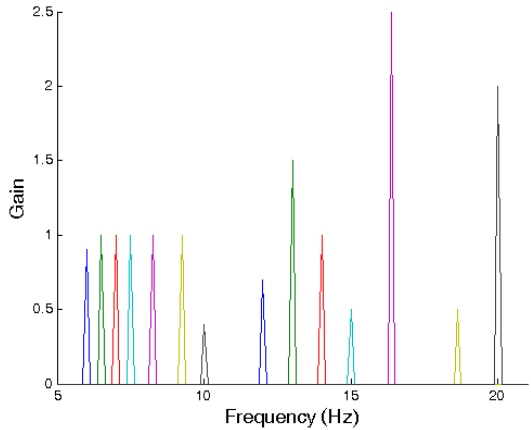

Fig. 3: Sample BIFB design for AVI Dataset.

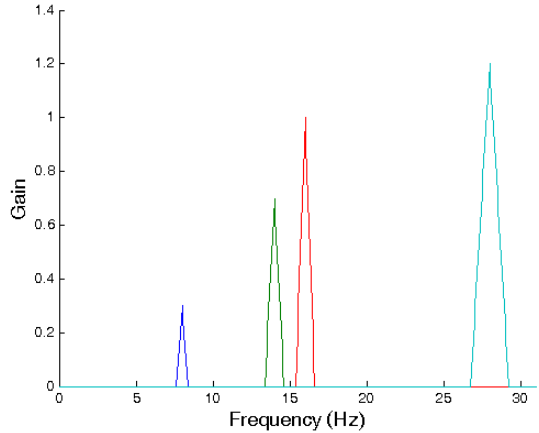

Fig. 4: Sample BIFB design for RIKEN-LABSP Dataset.

The second comparison method, CCA is a multivariable statistical method which is first used in [10] for SSVEP frequency detection. In summary, if $X$ is a multichannel EEG signal, $Y$ is the "Fourier series" of a simulated stimulus signal, and $w$ is linear combination coefficient CCA searches for the linear combination that maximizes correlation between $U = w_x^T X$ and $V = w_y^T Y$ by optimizing the following equation:

$$\max_{w_x w_y} \rho = \frac{E[w_x^T X Y^T w_y]}{\sqrt{E[w_x^T X X^T w_x] E[w_y^T Y Y^T w_y]}}$$

Only the highest correlation of canonical variables U and V are used for the frequency detection and this step is repeated for each simulated stimulus signal's Fourier series, Y. The stimuli frequency which provides the highest correlation among other stimuli frequencies is detected as the actual SSVEP frequency. Similar to PSDA and BIFB, the same EEG preprocessing steps are applied for CCA method as well.

### III. RESULTS

The results are evaluated with the most commonly used performance metrics for BCI algorithms [3]: accuracy, which indicates correct detection rate, and ITR which can be expressed as follows:

$$ITR = s \left[ \log_2(N) + p \log_2 p + (1-p) \log_2 \left( \frac{1-p}{N-1} \right) \right]$$

where $N$ stands for equiprobable user commands, $s$ denotes commands performed per minute, and $p$ represents the probability of correctly detected commands.

The proposed BIFB method for SSVEP frequency detection is tested on eight subjects and 151 trials in total using two datasets. The results which compare BIFB with the two baseline methods, PSDA and CCA, are listed in Table II & Table III. It should be pointed out that ITR changes logarithmically with the number of available commands and since numbers of commands between two datasets are different; ITRs need to be compared separately for each dataset.

The classical PSDA method needs longer time windows compared to other two methods to improve the accuracy which then leads to lower ITR. CCA, on the other hand, is successful on shorter time windows providing high ITR but lower accuracy. The accuracy can subsequently be improved by using longer time windows with lower ITR trade-off. Moreover, the performance of CCA in RIKEN LABSP dataset is poor since it is insufficient and incompatible to detect 28 Hz.

Both tables show that the BIFB method provides both reliable accuracy and sufficient ITR performance which is comparable with CCA due to its bio-inspired design. It is true that BIFB requires a longer training, or calibration process compared to CCA. However, the preliminary results show that even without any training, using a generic, non-user specific filter bank design, the accuracy values of BIFB are still comparable with CCA. As future work, this observation will be explored in detail to create a trade-off metric between training/accuracy/ITR trio.

### IV. CONCLUSION

A novel SSVEP detection method inspired by its biological characteristics is introduced in this paper. This method was shown to not only improve the accuracy but also increase the available number of commands by allowing the use of stimuli frequencies elicits weak SSVEP responses. The BIFB method achieved reliable performance metrics when tested on datasets available online and compared with two well-known SSVEP frequency detection methods as baseline methods: PSDA and CCA. The results show the potential of bio-inspired design which will be extended to include further SSVEP characteristics (e.g. time-domain waveform) and trade-off parameters to link accuracy and ITR with the level of training.

TABLE II. PERFORMANCE COMPARISON USING AVI SSVEP DATASET

| AVI SSVEP Dataset <br> # of Commands= 7 [6Hz, 6.5Hz, 7Hz, 7.5Hz, 8.2Hz, 9.3Hz, 10Hz], MDT= Mean Detection Time ||||||||||
|---|---|---|---|---|---|---|---|---|---|
| Subject | # of Trials | PSDA ||| CCA ||| BIFB |||
| | | MDT (sec) | Acc. (%) | ITR (bits/min) | MDT (sec) | Acc. (%) | ITR (bits/min) | MDT (sec) | Acc. (%) | ITR (bits/min) |
| Subject #1 | 24 | 10 | 83.3 | 10.4 | 5.5 | 87.5 | 21.2 | 7.4 | 91.7 | 17.7 |
| Subject #2 | 26 | 12 | 80.8 | 8 | 3.5 | 80.8 | 27.5 | 8.2 | 100 | 20.7 |
| Subject #3 | 21 | 10 | 85.7 | 11.1 | 4 | 90.5 | 31.6 | 7.4 | 100 | 22.7 |
| Subject #4 | 21 | 8 | 85.7 | 13.8 | 7 | 100 | 24.1 | 6.3 | 100 | 26.6 |

TABLE III. PERFORMANCE COMPARISON USING RIKEN-LABSP SSVEP DATASET

| RIKEN LABSP SSVEP Dataset <br> # of Commands= 3 [8Hz, 14Hz, 28Hz], MDT= Mean Detection Time ||||||||||
|---|---|---|---|---|---|---|---|---|---|
| Subject | # of Trials | PSDA ||| CCA ||| BIFB |||
| | | MDT (sec) | Acc. (%) | ITR (bits/min) | MDT (sec) | Acc. (%) | ITR (bits/min) | MDT (sec) | Acc. (%) | ITR (bits/min) |
| Subject #1 | 15 | 10 | 66.7 | 2.00 | 4 | 73.3 | 7.2 | 7.5 | 100 | 12.7 |
| Subject #2 | 15 | 9 | 66.7 | 2.22 | 4 | 60 | 3.2 | 7.8 | 100 | 12.2 |
| Subject #3 | 15 | 15 | 60 | 0.9 | 5 | 66.7 | 4 | 10 | 86.7 | 5.3 |
| Subject #4 | 15 | 15 | 6.7 | - | 3 | 66.7 | 6.67 | 8.33 | 66.7 | 2.4 |